\documentclass[11pt,preprint]{aastex}

\usepackage{amsmath}
\usepackage{graphicx}

\shorttitle{Contracting Universe}
\shortauthors{Sumner}

\begin{document}

\title{Observational Evidence from Supernovae for a Contracting Universe}

\author{William Q. Sumner\altaffilmark{1}}

\affil{Computer Science Department, Central Washington University, Ellensburg, WA 98926}

\altaffiltext{1}{sumner@cwu.edu}

\begin{abstract}

New precision in measuring extragalactic distances using supernovae has confirmed with high probability an accelerating increase in redshift with distance.  This has been interpreted as implying the existence of dark energy in an expanding and accelerating, flat universe.  A more logical explanation of these observations follows directly from an observation made by Erwin Schr\"odinger in 1939 that in a closed Friedmann universe $every$ quantum wave function changes with spacetime geometry.  Double the size of the universe and both the wavelengths of photons and the sizes of atoms double.  When the evolution of atoms and photons are combined, the meaning of Hubble redshift is reversed.  Redshift is characteristic of contracting universes.  The magnitude-redshift curve for a contracting universe has exactly the accelerating form recently observed and is in excellent quantitative agreement with the data of Riess et al. 1998, Knop et al. 2003, and others.  An observed maximum redshift of 1.3  gives a minimum age estimate for the universe of $114$ billion years.  The time until collapse is estimated to be 15 billion years or less. 

\end{abstract}


\section{Introduction}

Supernovae provide an excellent opportunity for measuring extragalactic distances. The observations of these exploding stars using the best earth and space-based telescopes, a physical understanding of supernovae evolution, and sophisticated analyses of their spectra have led to the surprising conclusion that the expansion of the universe seems to be accelerating (Perlmutter \& Schmidt 2003).  The upward trend of the magnitude versus redshift curve at high redshift is clear in Figure 1 (Riess et al. 1998). 

This departure from a linear or decelerating trend has been explained as eternal expansion with a positive cosmological constant and a current acceleration of expansion.  This conclusion is based on a conventional interpretation of redshift as the result of photon wavelength increasing in an expanding Friedmann universe.  This was one idea originally suggested by Hubble \& Tolman (1935), but it is interesting to note that they concluded their paper with

\begin{quote}
It also seemed desirable to express an open-minded position as to the true cause of the nebular red-shift, and to point out the indications that spatial curvature may have to play a part in the explanation of existing nebular data.
\end{quote}

The conclusion drawn here is that spatial curvature indeed plays a critical part in explaining cosmic redshift.  It is the combination of changes in photons during their transit time and changes in atoms between the time of photon emission and later when they provide the wavelength standard, that defines redshift.    

\section{Redshift}

The Friedmann (1922) line element for the closed solution to Einstein's (1916) theory of general relativity with no cosmological constant may be written 

\begin{equation}
ds^2\,=\,c^2dt^2\,-\,a^2 (t)\left[\frac{dr^2}{\left(1-r^2\right)}\,+\,r^2\left(d\vartheta^2\,+\,sin^2\vartheta \,d\varphi^2\right)\right]. 
\end{equation}

Combined with electrodynamics, this gives the well-known result that photon wavelength $\lambda(t)$ evolves proportionally to the Friedmann radius $a(t)$. In an expanding universe, photon wavelengths increase in time and in a contracting universe, photon wavelengths decrease,

\begin{equation}
\frac{\lambda (t_0)}{\lambda (t_1)}\,=\,\frac{a(t_0)}{a(t_1)}. 
\end{equation}

Schr\"odinger (1939) also derived this equation as an example of the coevolution of quantum wave functions with the curved spacetime of the closed Friedmann universe.  Schr\"odinger found that the plane-wave eigenfunctions characteristic of flat spacetime are replaced in the curved spacetime of the Friedmann universe by wave functions that are not precisely flat and that have wavelengths that are directly proportional to the Friedmann radius $a(t)$. This means that eigenfunctions change wavelength as the radius of the universe changes and every quantum system they describe changes. In an expanding universe quantum systems expand. In a contracting universe they contract.

An equation similar to (2) also describes the evolution of other quantum systems such as atoms, where $\lambda (t)$ is a measure of atomic size (Schr\"odinger 1939, Sumner \& Sumner 2000).  Change in atomic size is directly related to change in the wavelengths of light that atoms emit.  The shift in wavelength of a characteristic atomic emission $\lambda_e (t)$ due to coevolution is (Sumner 1994)  

\begin{equation}
\frac{\lambda_e (t_0)}{\lambda_e (t_1)}\,=\,\frac{a^2(t_0)}{a^2(t_1)}. 
\end{equation}  

Taken together, equations (2) and (3) imply that redshift is observed only when the Friedmann Universe is contracting (Sumner 1994).  Blueshifts are expected in an expanding universe.  

Redshift $z$ is traditionally defined 

\begin{equation}
z\,=\,\frac{\lambda_{obs} (t_0)\,-\,\lambda}{\lambda}, 
\end{equation} 

\noindent where $\lambda$ is the assumed constant atomic emission and $\lambda_{obs}(t_0)$ is the photon wavelength observed. Making this traditional assumption that only photons evolve gives

 \begin{equation}
 z\,=\,\frac{a(t_0)}{a(t_1)}\,-\,1, 
 \end{equation} 
 
\noindent where $a(t_0)$ is the current radius and $a(t_1)$ was the radius at the time of emission.  This definition must be reconsidered to account for the evolution of atomic emissions in addition to photon evolution. 

Define redshift $k$ as 

\begin{equation}
k\,=\,\frac{\lambda_{obs} (t_0)\,-\,\lambda_e(t_0)}{\lambda_e(t_0)}, 
\end{equation} 

\noindent where $\lambda_e(t_0)$ is the wavelength emitted by today's reference atom and $\lambda_{obs}(t_0)$ is the wavelength observed today from a distant source.  Experimental measurements actually determine the redshift $k$, not $z$.  Equation (6) is equivalent to 

\begin{equation}
k\,=\,\frac{a(t_1)}{a(t_0)}\,-\,1. 
\end{equation} 

The mathematical coordinate distance $r_1$ is directly related to the observed redshift $k$ of a source and the deceleration parameter $q_o$.  The derivation parallels the one done when atomic evolution is ignored and the universe is assumed to be expanding (Narlikar 1993), but differs because $k$, not $z$, describes the observed redshift and because some sign choices must be made differently when the cycloid parameter $\theta$ is in the third and fourth quadrants.   If it is assumed that the observed photons were emitted after contraction began, $r_1$ is given by (Sumner \& Vityaev 2000)

\begin{equation}
r_1= \frac{\left(2q_o-1\right)^{1/2}}{q_o}\left[k-\frac{(1+k)(1- q_o)}{q_o}\right] + \frac{(1-q_o)}{q_o}\left\{ 1-\left[k-\frac{(1+k)(1- q_o)}{q_o}\right]^2\right\}^{1/2}  .
\end{equation}

Luminosity distance $D_L$ is related to the observed flux $f$ and the intrinsic luminosity $L$ of the source by the equation \begin{equation}f\,=\,\frac{L}{4\pi D_L^2}. \end{equation}

The flux observed can be calculated by noting that the luminosity is decreased by a factor of $a(t_0)/a(t_1)$ due to the apparent decrease of the photon's energy and decreased by another factor of $a(t_0)/a(t_1)$ due to the changes in local time.  The distance to the source is $r_1\,a(t_0)$.  This gives an observed flux of 

\begin{equation}
f\,=\,\frac{L\frac{a^2(t_0)}{a^2(t_1)}}{4\pi r_1^{2}a^2(t_0)}. 
\end{equation}

Comparing (9) and (10) gives 

\begin{equation}
D_L\,=\,r_1\,a(t_0)\,(1\,+\,k), 
\end{equation}

\noindent where equation (7) has been used.  Combining equations (8) and (11) and noting that  

\begin{equation}
a(t_0)=\frac{-c}{H_o(2q_o\,-\,1)^{1/2}}, 
\end{equation} 

\noindent gives the desired result 

\begin{equation}
\begin{split}
D_L\,=&\,\left(\frac{-c}{H_o}\right) \frac{(1+k)}{q_o}\Biggl\{\left[k-\frac{(1-k)(1- q_o)}{q_o}\right]+  \\
&+\frac{(1-q_o)}{(2q_o-1)^{1/2}}\left( 1-\left[k-\frac{(1+k)(1- q_o)}{q_o}\right]^2\right)^{1/2}\Biggr\}. \\
\end{split}
\end{equation}

Equations (12) and (13) contain minus signs since $H_o$ is negative for contracting universes.

\section{Redshift Observations}

Recent determinations of the relationship between magnitude and redshift using supernovae have achieved an accuracy that can distinguish between various theoretical possibilities.  The most important conclusion is that a conventional interpretation of redshift using an expanding Friedmann metric and no cosmological constant is inconsistent with the data. Figure 1 nicely illustrates this conclusion.

Taking this same data set one, can use equation (13) and vary $H_o$ and $q_o$ for a best fit. Figure 2 illustrates a fit for the parameters $H_o\,=\,-65.1\, km\,s^{-1}\,Mpc^{-1}$ and $q_o\,=\,0.51$.  The relationship between magnitude and luminosity distance 

\begin{equation} 
m\,-\,M\,=\,5\,\log_{10}\left(\frac{D_L}{10\,parsecs}\right), 
\end{equation} 

\noindent was used.

Figure 3 illustrates a similar fit for the most robust data set from Knop et al. 2003, with values of $H_o\,=\,-69.7\, km\,s^{-1}\,Mpc^{-1}$ and $q_o\,=\,0.51$.  A value of 19.27 was added to their values for $m_B^{eff}$ which corresponds to using a value of $H_o\,=\,-70\, km\,s^{-1}\,Mpc^{-1}$ to set the distance scale. This choice determines the best fit $H_o$ but has no other affect.  For example, if 19.41 is added (corresponding to $H_o\,=\,-65\, km\,s^{-1}\,Mpc^{-1}$) the best fit is $H_o\,=\,-65.2\, km\,s^{-1}\,Mpc^{-1}$ and $q_o\,=\,0.51$, with no difference in the standard error.  

Standard error, $S_y$, is defined in the usual way as the root-mean-square of the differences, $d_i$, between $n$ data and the value of equation (13),

\begin{equation}
S_y= \sqrt {\Sigma d_i^{\,2}/n}.
\end{equation}

The standard errors for both of these choices for $H_o$ with Knop's data and for the fit to the data of Riess have exactly the same value $0.188$.  This is comparable to the stated data errors.  Figure 4 illustrates the fit sensitivity to parameter choice. 

The high redshift supernovae data of Tonry et al. (2003) and Barris et al. (2003) are entirely consistent with these results.   Analyzed alone they are easy to fit, but neither $H_o$ nor $q_o$ are tightly constrained.  When combined with lower redshift data, they have similar fits to the ones shown here.

\section{Age}

The age of the universe may be estimated from the magnitude-redshift data in the usual manner (Narlikar 1993), except one sign must be changed to reflect contraction,  

\begin{equation}
t_o= \frac{-1}{H_o}\left[\frac{1}{2q_o-1}+\frac{q_o}{\left(2q_o-1\right)^{3/2}}\cos^{-1}{\frac{1-q_o}{q_o}} \right] .
\end{equation}

\noindent A value for $\cos^{-1}$ corresponding to the third or fourth quadrant must also be used.

A method to give a robust estimate for a minimum age is to note that for a given maximum value of redshift, $k_{max}$, a minimum age is defined by the Hubble constant $H_o$ and by the condition that $q_o$ be maximum.  In equation (13), this maximal condition occurs when the term in the square root is zero,

\begin{equation}
\left( 1-\left[k_{max}-\frac{(1+k_{max})(1- q_o)}{q_o}\right]^2\right)^{1/2} = 0,
\end{equation}

\noindent which is equivalent to

\begin{equation}
q_o=\frac{1+k_{max}}{2k_{max}}.
\end{equation}

The redshift of 1.3 (Riess et al. 2003), gives $q_o=0.88$.  With $H_o\,=\,-65\, km\,s^{-1}\,Mpc^{-1}$, equation (16) gives a minimum age estimate of 114 billion years.  

$ | H_o | ^{-1}$ provides an estimate for the time remaining until collapse.  This time is 15 billion years for $H_o =-65\, km\,s^{-1}\,Mpc^{-1}$.  This estimate is good for $q_o$ close to $0.5$ but  decreases to about 8 billion years when $q_o$ is close to $1$.

\section{Conclusion}

Recent supernovae data may be interpreted as providing experimental confirmation of the theoretical predictions of the coevolution of spacetime geometry and quantum wave functions made by Schr\"odinger (1939), Sumner (1994), and Sumner \& Sumner (2000).  This explanation of the magnitude-redshift relationship comes directly and unambiguously from the roots of the best-tested physical theories of the 1930's.

\bigskip 
\medskip 

The author thanks Dawn Sumner for many insights and helpful suggestions.

% Figures

\begin{figure}[t]
\centering
\centering \includegraphics[width=15cm]{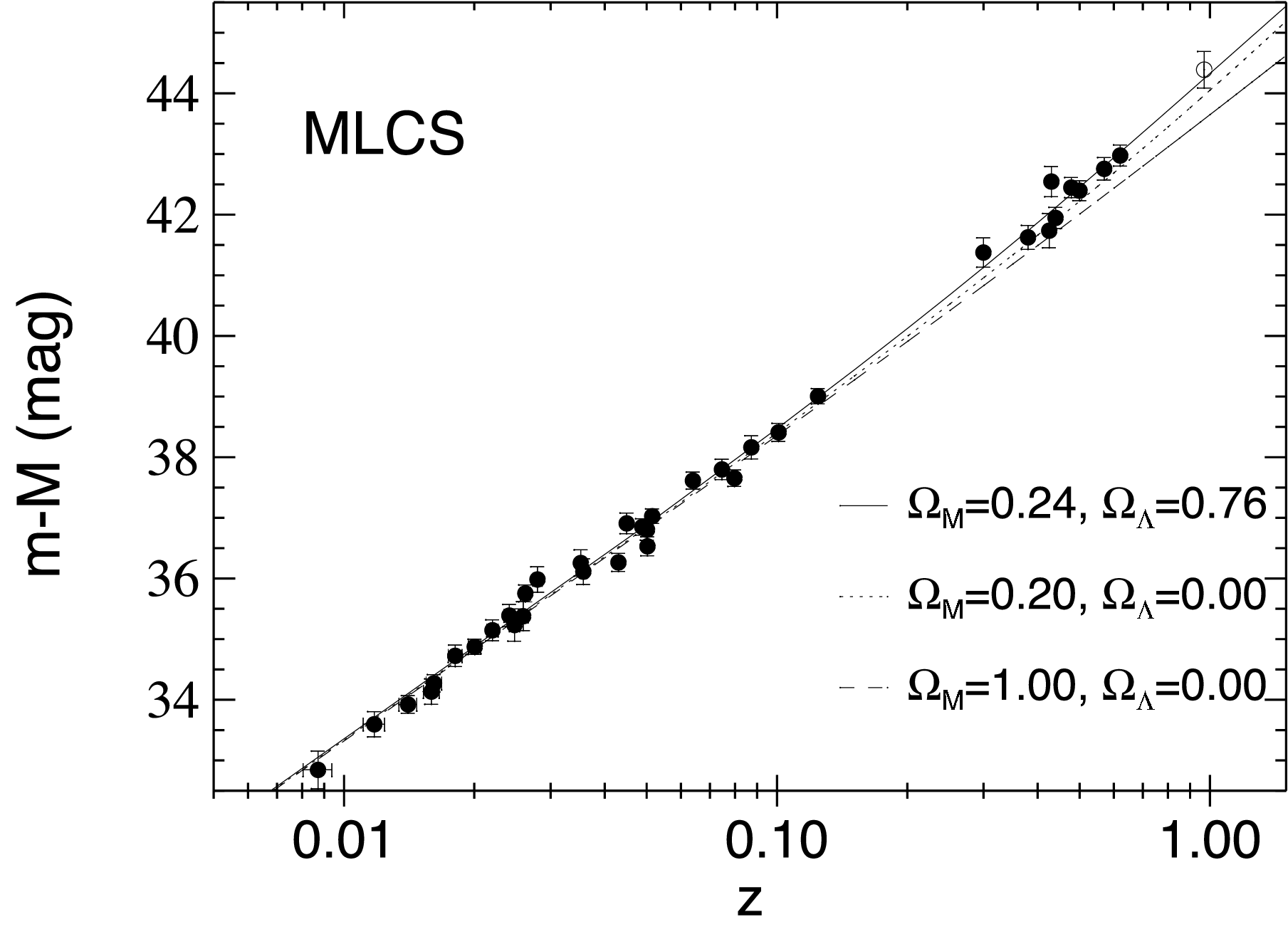}
\caption{Redshifts and magnitudes for 37 supernovae and best fits for three mixes of ordinary matter ($\Omega_M$) and vacuum energy resulting from a cosmological constant ($\Omega_\Lambda$).  Taken from Figure 4, Riess et al. 1998.}
\end{figure}

\begin{figure}[t]
\centering
\centering \includegraphics[width=15cm]{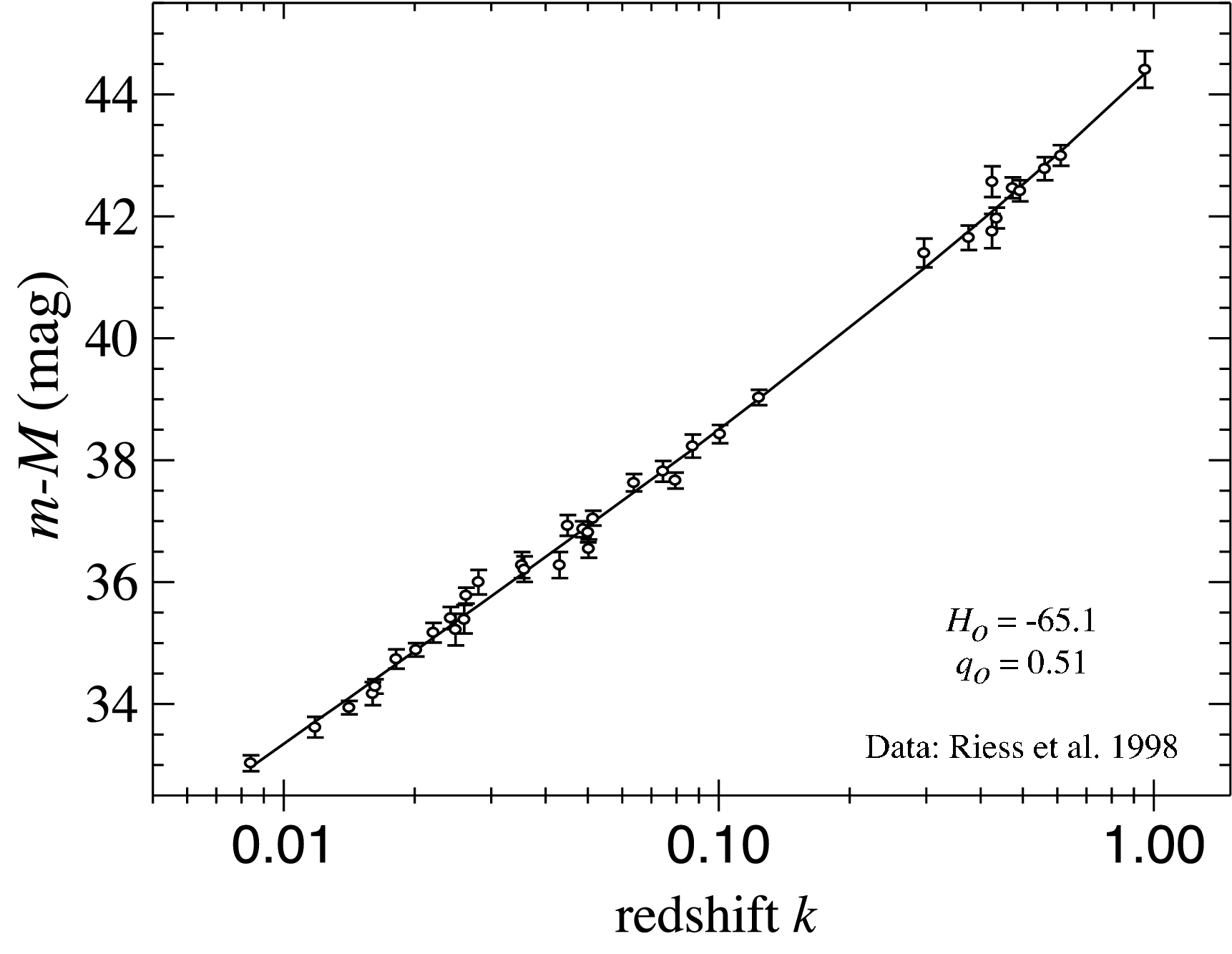}
\caption{Redshifts and magnitudes for 37 supernovae (Riess et al. 1998) and a fit with the parameters $H_o\,=\,-65.1\, km\,s^{-1}\,Mpc^{-1}$ and $q_o\,=\,0.51$ in equation (13).}
\end{figure}

\begin{figure}[t]
\centering
\centering \includegraphics[width=15cm]{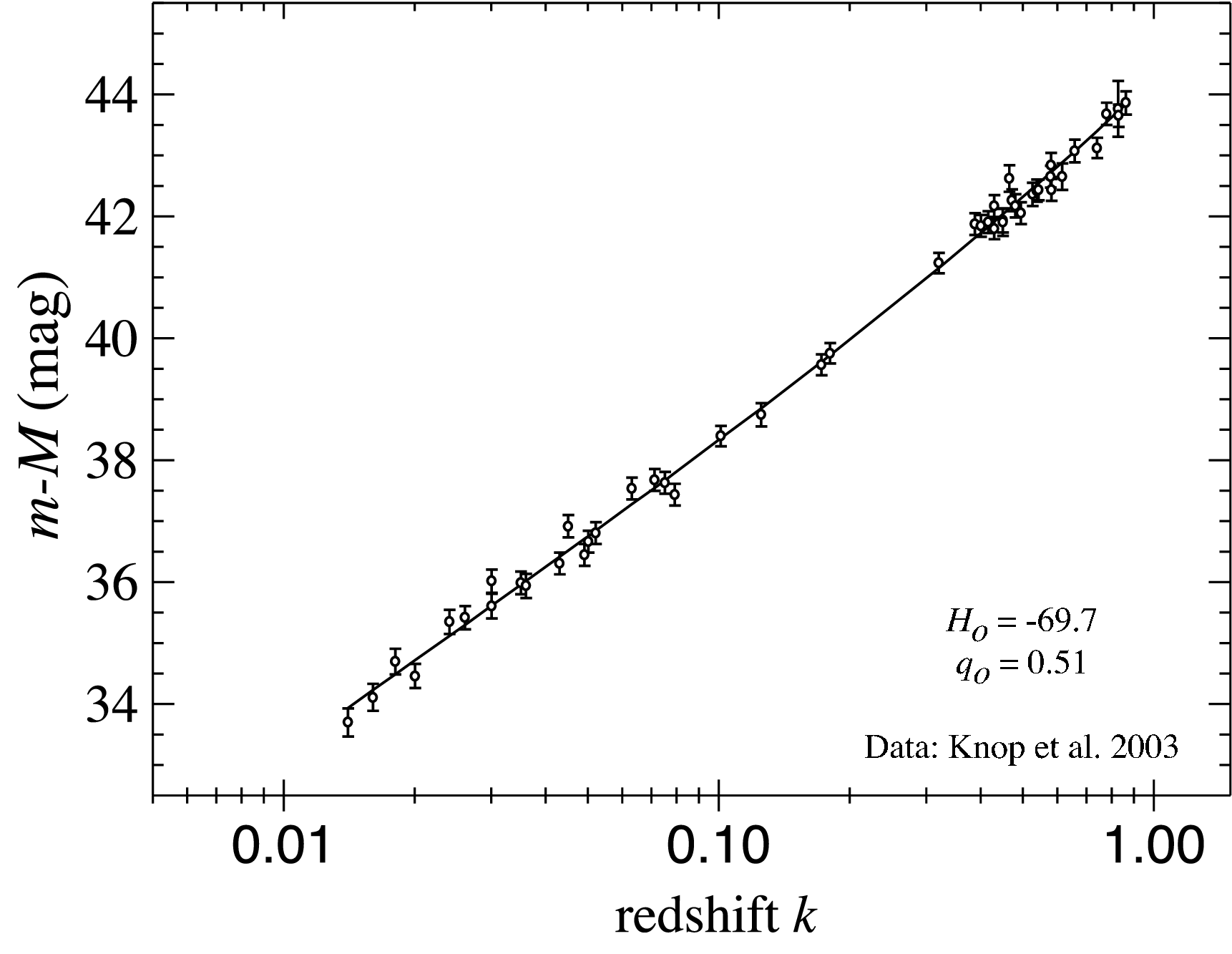}
\caption{Redshifts and magnitudes for 48 supernovae (Knop et al. 2003) and a fit with the parameters $H_o\,=\,-69.7\, km\,s^{-1}\,Mpc^{-1}$ and $q_o\,=\,0.51$ in equation (13).}
\end{figure}

\begin{figure}[t]
\centering
\centering \includegraphics[width=12cm]{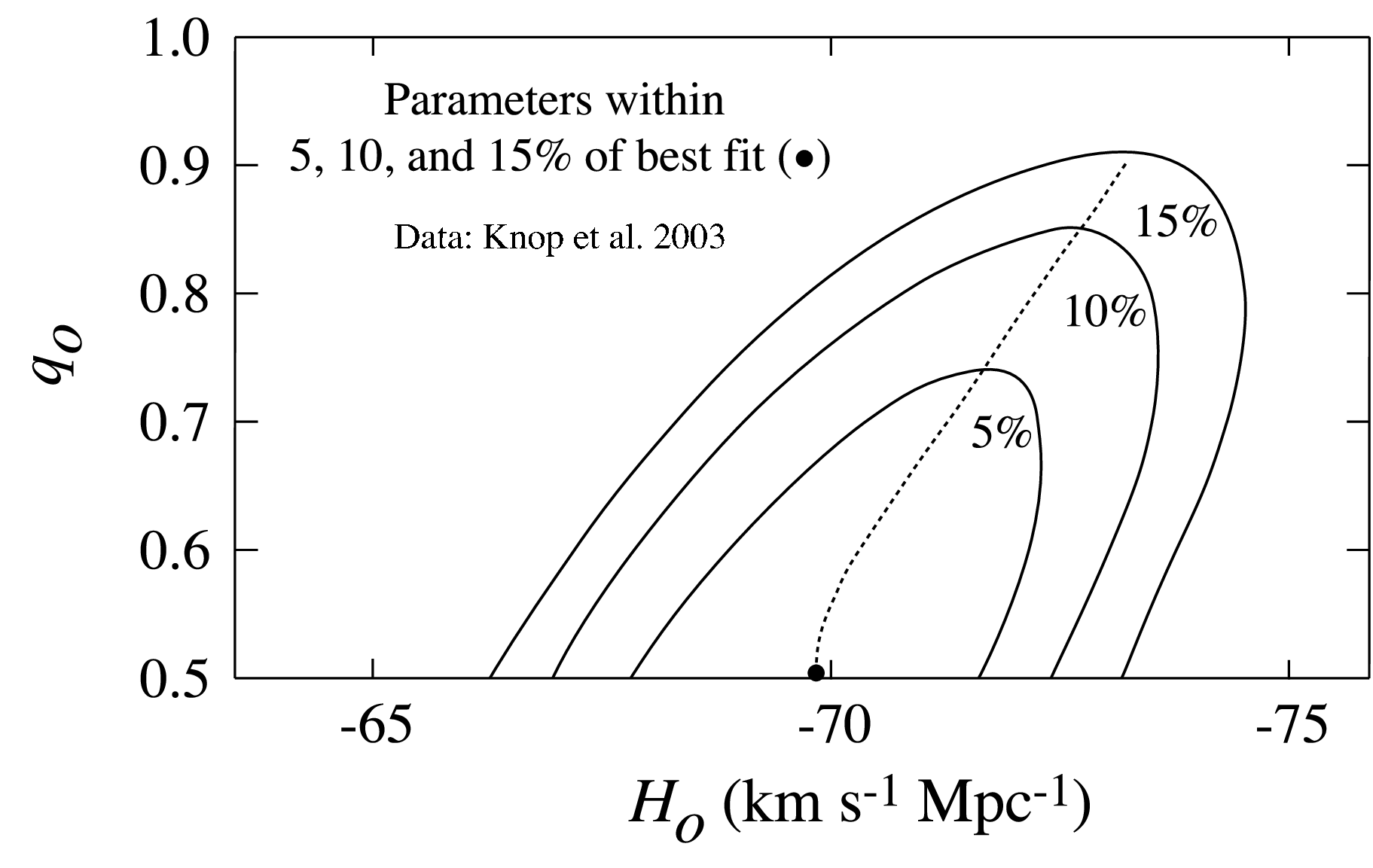}
\caption{Parameters within 5, 10, and 15\% of the best fit  $H_o\,=\,-69.7\, km\,s^{-1}\,Mpc^{-1}$ and $q_o=0.5+|\epsilon|$, where $\epsilon \approx 0$, for the data of Knop et al. 2003.  The dotted line indicates the minimum standard error for a given $q_o$ or for $H_o\,>\,-69.7\, km\,s^{-1}\,Mpc^{-1}$.  For $H_o\,<\,-69.7\, km\,s^{-1}\,Mpc^{-1}$, the minimum standard error is when $q_o=0.5+|\epsilon|$, where $\epsilon \approx 0$.}
\end{figure}

\end{document}